\documentclass[reprint,
 superscriptaddress,
 amsmath,amssymb,
 aps,
 prl,
]{revtex4-2}

	\usepackage[T1]{fontenc}
	\usepackage[utf8]{inputenc}
    \usepackage[pdftex,bookmarks=false]{hyperref}
    \usepackage{microtype}
    \usepackage{amssymb, amsmath}
    \usepackage{graphicx}
    \usepackage{dcolumn}
    \usepackage{bm}
    \usepackage{multirow}
    \usepackage{booktabs}
    \usepackage{xcolor}
    \usepackage{mathtools}
    \usepackage{mathptmx}
    \usepackage{upgreek}
    \usepackage{float}
    \usepackage{lipsum}
    \usepackage{siunitx}
    \usepackage{array}
    \usepackage{enumitem}
    \usepackage[inkscapelatex=false]{svg}
    \usepackage{makecell}
    
    \hypersetup
    {
    	colorlinks = true, 
		citecolor = blue,
		bookmarksnumbered = true,
		urlcolor = green,
		pdfauthor = {},
		pdftitle = {},
		pdfsubject = {},
		pdfkeywords = {}
	}
	
    \DeclarePairedDelimiterX\braket[2]{\langle}{\rangle}{#1 \delimsize\vert #2}

    \graphicspath{Figures}

\newcolumntype{C}{>{\centering\arraybackslash}m{0.75cm}}

\begin{document}

\title{
    Active learning potentials for first-principles phase diagrams using replica-exchange nested sampling
}

\date{\today}
\author{N.~Unglert}
\affiliation{Institute of Materials Chemistry, TU Wien, 1060 Vienna, Austria} 

\author{M.~Ketter}
\affiliation{Institute of Materials Chemistry, TU Wien, 1060 Vienna, Austria} 

\author{G. K. H. Madsen}
\email[Correspondence email address: ]{georg.madsen@tuwien.ac.at}%
\affiliation{Institute of Materials Chemistry, TU Wien, 1060 Vienna, Austria}

\begin{abstract}
    Accurate prediction of materials phase diagrams from first principles remains a central challenge in computational materials science. 
    Machine-learning interatomic potentials can provide near-DFT accuracy at a fraction of the cost, 
    but their reliability crucially depends on the availability of representative training data that span all relevant regions of the potential-energy surface. 
    Here, we present a fully automated active-learning (AL) strategy based on replica-exchange nested sampling (RENS) for the generation of training data and the computation of complete pressure–temperature phase diagrams. 
    In our framework, RENS acts as both the exploration engine and the acquisition mechanism: 
    its intrinsic diversity and likelihood-constrained sampling ensure that the configurations selected for DFT labeling are both informative and thermodynamically representative. 
    We apply the approach to silicon, germanium, and titanium using potentials trained at the r2SCAN level of theory. 
    For all systems, the AL process converges within $\sim$10–15 iterations, yielding transferable potentials that reproduce known phase transitions and thermodynamic trends. 
    These results demonstrate that RENS-based AL provides a general and autonomous route to constructing machine-learning interatomic potentials  and predicting first-principles phase diagrams across broad thermodynamic conditions.
\end{abstract}

\maketitle

\section{Introduction}
Predicting phase diagrams from first principles remains a long-standing challenge in materials science. 
Accurate modeling of phase stability requires access to the free energy across vast regions of configuration space, where competing structural motifs and bonding types coexist. Nested sampling (NS) offers a statistical-mechanics framework for addressing this challenge. 
Originally developed as a Bayesian inference algorithm \cite{skilling_nested_2004}, NS provides direct access to the configurational partition function and thus to thermodynamic observables over the entire temperature range from a single simulation. 
In materials science, NS has proven particularly effective for mapping complex phase diagrams and exploring multiphase energy landscapes \cite{partay_nested_2021}.

Accurate phase diagrams, however, presuppose a quantitatively reliable description of the potential-energy surface (PES). While density functional theory (DFT) can provide such accuracy, its high computational cost restricts the accessible regions of phase space. 
Machine-learning interatomic potentials (MLIPs) have emerged as a powerful means to overcome these limitations by reproducing DFT-level accuracy at a fraction of the computational cost \cite{deringer_machine_2019,leimeroth_machine-learning_2025} and have enabled atomistic simulations of complex materials ranging from covalent solids and metallic alloys to liquids and amorphous phases \cite{erhard_modelling_2024, shenoy_collinear-spin_2024,menon_electrons_2024,erhard_machine-learned_2022}.
However, constructing the underlying training databases remains a major bottleneck to the autonomous use of MLIPs in thermodynamic simulations. Generating representative and diverse reference data in a closed loop remains challenging. As a result, combinations of MLIPs with NS, whose systematic PES exploration spans regions of high energy and low probability, have typically relied on carefully handcrafted datasets \cite{unglert_neural-network_2023,kloppenburg_general-purpose_2023,marchant_exploring_2023}.

Active learning (AL) provides a principled route to eliminate manual data generation. 
In AL, the model iteratively selects configurations and retrains itself with the augmented dataset. 
A wide range of AL schemes have been explored in atomistic modeling \cite{bernstein_novo_2019,jinnouchi_phase_2019,zaverkin_exploring_2022,Wanzenboeck_DD24,schafer_apax_2025},
but their efficiency depends critically on how the exploration and acquisition stages are coupled to the underlying sampling algorithm. Methods like NS evolve through a sequence of changing target distributions, and therefore require a dedicated strategy to incorporate AL feedback efficiently \cite{Fletcher_arXiv25}.

In this work, we combine the strengths of AL and NS in an automated framework in which the sampling, labeling, and retraining stages are seamlessly integrated. 
A key component of our approach is the use of replica-exchange nested sampling (RENS) \cite{unglert_replica_2025}, which incorporates swap moves reminiscent of parallel tempering.
This substantially improves ergodicity by allowing replicas to swap configurations and thus explore neighbouring regions of the enthalpy landscape. For many systems, this not only enhances sampling efficiency but renders the exploration of complex phase behaviour feasible in the first place \cite{unglert_replica_2025}. 

To demonstrate the approach, we apply our RENS based AL strategy to three elemental systems: silicon, germanium, and titanium. 
Silicon serves as a well-established benchmark, for which high-quality reference data and prior NS results are available \cite{bartok_machine_2018,unglert_neural-network_2023,unglert_replica_2025}.
Germanium provides an electronically related test case, while titanium represents a distinct metallic system. We discuss how RENS yields configurations that are both physically relevant and broadly distributed across thermodynamic states, eliminating the need for explicit distance metrics or clustering criteria to enforce dataset diversity.
Together, the examples showcase the fully automated construction of MLIPs based on configurations actively selected during the RENS algorithm and the direct computation of first-principles pressure–temperature phase diagrams across broad thermodynamic ranges.

\section{Results}

\subsection{RENS active learning}

In atomistic simulations, an AL strategy is typically implemented as an iterative process comprising three key steps: (i) a MLIP-based exploration algorithm generates atomic configurations and incorporates an acquisition mechanism that selects configurations to be added to the training database, (ii) a ground-truth evaluation procedure provides reference labels (e.g. DFT energies and forces) for the selected configurations and (iii) the MLIP is retrained using the updated dataset.
The design of an AL strategy depends on the computational cost associated with each of these steps. 

For models such as Gaussian process regression, retraining is computationally inexpensive, making it feasible to update the model after each newly selected data point. In contrast, training modern neural-network-based MLIPs is substantially more demanding. As a result, batch-mode strategies, where retraining is triggered only after a batch of new data has been accumulated, are typically more practical.

\begin{figure}
    \centering
    \includegraphics[width=.8\columnwidth]{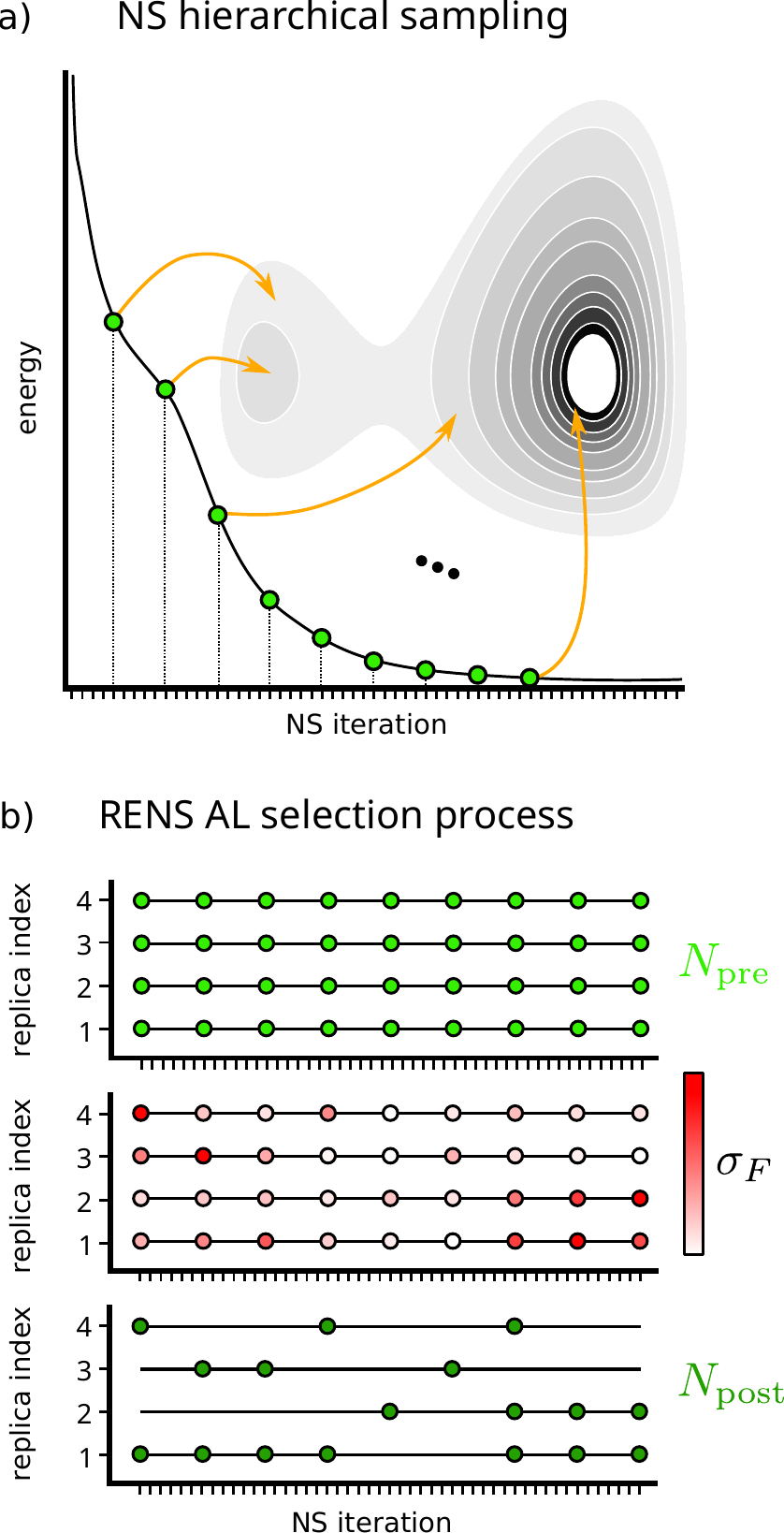}
    \caption{
        a) Schematic plot of a NS simulation, indicating the monotonically decreasing energy of the NS samples as well as a sequence of regularly spaced samples ${R_i}$ as green dots together with the hypershells in configuration space that can be assigned to them.
        b) Schematic plot of a $M=4$ RENS simulation, indicating the sample trajectories as black lines and the $N_\mathrm{pre}$ sliced out samples as green circles. Middle panel shows the latter colored according to their uncertainty $\sigma_F$ and bottom panel shows remaining batch of $N_\mathrm{post}$ samples after uncertainty subsampling.
    }
    \label{fig:al_strategies}
\end{figure}

We deliberately restrict ourselves to a naive batch-mode strategy, where the sampling algorithm is executed to completion, producing a full configuration sequence from which a batch of data points is selected afterwards. 
In a RENS-based AL framework, this approach allows the progress of the AL loop to be monitored at each iteration by evaluating the full phase diagram corresponding to the current model state.
We apply a schedule to the parameters of the RENS algorithm. This is particularly useful for expensive exploration techniques that involve cost–performance tradeoffs. 
NS evolves a population of $K$ walker configurations of size $N_\mathrm{atoms}$ through random walks of length $L$.
In each iteration, the configuration with the lowest likelihood is removed and recorded as a sample that can be associated with a hypershell of decreasing configuration-space volume,  Fig.~\ref{fig:al_strategies}a.
RENS furthermore introduces an exchange mechanism that couples $M$ replicas, in this case each carrying out an independent NS run at different external pressures. In RENS, such tunable parameters therefore include the number of replicas $M$, the number of walkers $K$ and the walk length $L$.

Each of the $M$ replicas encodes a nested trajectory of length $N_\mathrm{iter}$. As illustrated in Fig.~\ref{fig:al_strategies}b (top), we extract a preliminary set of $N_\mathrm{pre}$ configurations by slicing these trajectories at regular intervals along the iteration index.  
For each configuration, uncertainty metrics are evaluated according to Ref.~\cite{heid_spatially_2024}, where committee standard deviations serve as estimators of model uncertainty. The energy uncertainty is given as 
\begin{align*}
    \label{eq:energy_uncty}
    \sigma_E &= \sqrt{\frac{1}{N_C} \sum_{l=1}^{N_C} 
    \left(E^{(l)} - \overline{E}\right)^2}.
\end{align*}
where $\overline{E}$ is the mean of the individual energies $E^{(l)}$ of all $N_C$ committee members.
For the force-based estimator, we employ a slightly modified variant, where we compute the standard deviation of the force component $f_i^k$ of atom $i$ and spatial direction $k$ via
\begin{align}
    s^{(k)}_i = \sqrt{
        \frac{1}{N_C} \sum_l^{N_C} 
        \Big( f^{(kl)}_i - \overline{f}^{(k)}_i \Big)^2
    },
\end{align}
where $f^{(kl)}_i$ are the predicted force components for committee member $l$ and $\overline{f}^{(k)}_i = N_C^{-1} \sum_l  f^{(kl)}_i$.
We aggregate these standard deviations over the whole structure according to 
\begin{align}
    \label{eq:force_uncty}
    \sigma_F 
        = \Big(\sum_{ik} \overline{f}^{(k)}_i \Big)^{-1}
        \sum_{ik} s^{(k)}_i
    .
\end{align}
This uncertainty measure is normalized by the total magnitude of the mean force.
We base this normalization on the observation that the unnormalized force uncertainty tends to strongly favor high energy configurations. Hence, the normalization acts as a regularizer for the uncertainties in scenarios where configurations are sampled across multiple energy scales. 
A final subset of $N_\mathrm{post}$ configurations is resampled without replacement, using the uncertainty values as weights. This yields the final selection ${N_\mathrm{post}}$ configurations for ground-truth evaluation in the AL loop. This procedure avoids redundant selection of similar configurations and obviates the need for explicit distance-based diversity controls. Nevertheless, such metrics could be readily incorporated on top of the present scheme to further improve selection efficiency if desired.

The subsampling of the prelimenary RENS configurations is thus purely based on uncertainty. This favors informative configurations but can introduce redundancy or lead to poor coverage of configuration space \cite{zaverkin_exploring_2022}. 
We performed initial tests comparing random sampling from an NS run to sampling according to greedy distance maximization \cite{zaverkin_exploring_2022} using the average minimum distance metric \cite{widdowson2022average}. Both selection strategies yielded MLIPs of similar predictive quality for the relevant solid phases. We attribute this to the NS procedure itself. The hierarchical nature of NS naturally provides a diverse coverage of configuration space, Fig.~\ref{fig:al_strategies}a while the replica-exchange mechanism ensures ensemble connectivity across replicas, maintaining continuity of sampling even when the accessible configuration space shrinks rapidly during phase transitions \cite{unglert_replica_2025}, reducing the benefit of explicit diversity-driven selectors.

For all three cases, Si, Ge, and Ti, we constructed initial databases by extracting all experimentally reported entries of the target material from the materials project \cite{jain_commentary_2013, horton_accelerated_2025}. To enrich the structural diversity, each configuration was modified in three ways: (i) A supercell was created that fits a predefined cutoff (ii) The volumes were scaled isotropically using a predefined set of scaling factors and (iii) atomic displacements generated by adding independent Gaussian noise. The details of the initial databases are given in the Methods section.
In the following, we describe the AL process for each material and 
present accurate RENS phase diagrams obtained with the resulting AL MLIPs.

\subsection{Silicon}

\begin{figure}
    \centering
    \includegraphics[width=1.\columnwidth]{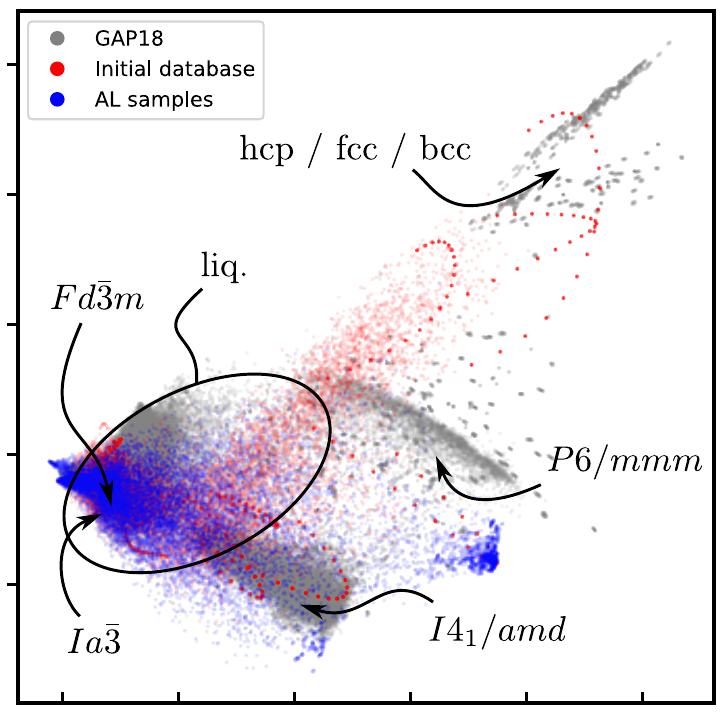}
    \caption{
        Comparison of atomic environments for all configurations contained in the GAP-18 silicon database (grey points) and the configurations used as initial database for the silicon AL run (red points). All configurations collected by the AL run are superimposed as blue points.
    }
    \label{fig:si_initial_db}
\end{figure}
The initial database for the silicon comprises 480 configurations covering different crystalline phases.
To analyze its coverage, we represent the configurations using the same spherical Bessel descriptors used for our MLIP. As a reference, we do the same for the GAP-18 silicon database \cite{bartok_machine_2018}. This database contains around 2475 manually curated silicon structures and has been shown to yield accurate silicon phase diagrams when combined with RENS \cite{unglert_replica_2025}. We perform a two-dimensional principal component analysis (PCA) on the GAP-18 database and show the projection of both the GAP-18 and initial database atomic environments onto these PCs in Figure~\ref{fig:si_initial_db}. For orientation, regions corresponding to prominent phases are highlighted. The comparison indicates that the initial database already spans a substantial fraction of the region covered by the GAP-18 database. However, the solid phases are populated much sparser in the initial database and the structures corresponding to disordered liquid phases are absent by construction.

\begin{figure*}
    \centering
    \includegraphics[width=1.\textwidth]{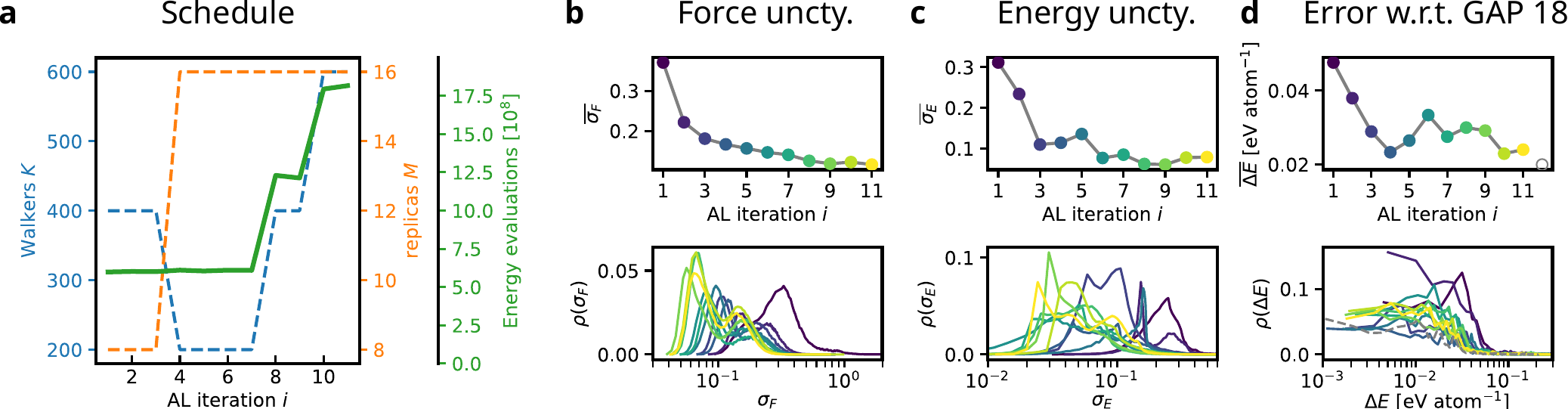}
    \caption{
        Schedule and metrics of a silicon AL run. 
        a) Schedule imposed on the $\mathrm{Si}$ sampling procedure, varying the number of walkers $K$ and the number of replicas $M$ together with an estimated number of energy evaluations each RENS simulation consumed. 
        b) and c) Force and energy uncertainties computed for all NS sample trajectories of AL iteration $i$. 
        d) Energy error on the GAP-18 silicon database for the model used for the RENS simulation in iteration $i$. The empty grey circle and dashed grey line indicate the performance of the model including the samples from the last AL iteration $i=11$.
        Bottom panels show the  distribution of the respective quantity, top panels the average values for each AL iteration. 
    }
    \label{fig:si_uncertainty}
\end{figure*}
The AL loop was initiated using the initial silicon database. In each AL iteration, the selection mechanism sliced out $N_\mathrm{pre}=600$ configurations from the RENS simulation and subsamples $N_\mathrm{post}=100$ configurations according to the normalized force-uncertainty. Since RENS simulations remain the computational bottleneck of the AL strategy, we vary the number of walkers $K$ and the number of replicas/pressures $M$ simulated following a predefined schedule, Fig.~\ref{fig:si_uncertainty}a. The choice of relatively inexpensive NS parameters allows the AL cycle to rapidly correct major deficiencies in the MLIP without expending large computational resources on accurately sampling an erroneous PES. As model uncertainty decreases, computational effort can be dedicated to finer sampling of the enthalpy landscapes. 
The number of energy evaluations of each RENS simulation was estimated by the product of the number of NS iterations, the MCMC walk length and the number of replicas, $N_\mathrm{iter} \cdot L \cdot M$. These numbers demonstrate the computational cost of RENS in the order of $10^9$ energy evaluations per run, Fig.~\ref{fig:si_uncertainty}a.
Although predefined in the present work, the scheduling could be automated in the future. Along similar lines, future work may also target an automated choice of replica pressure spacings that minimizes redundant sampling on-the-fly, e.g., by concentrating replicas around challenging phase boundaries. Such an approach would, however, require careful interpretation of the replica-exchange acceptance rates. A more detailed discussion of the role of this can be found in the SI. 

Figure~\ref{fig:si_uncertainty}b, c  summarize the progress of the AL procedure for silicon in terms of the evolution of uncertainty metrics, Eqs.~\eqref{eq:energy_uncty} and \eqref{eq:force_uncty}, where both the distribution and the average are depicted as function of the AL iteration. For the force uncertainties in Fig.~\ref{fig:si_uncertainty}b, a nearly monotonic reduction is observed. Interestingly, the distribution of the normalized force uncertainties shows a double-peaked structure in each run. With increasing AL iterations, both peaks shift systematically to lower magnitudes. A similar almost monotonic decrease is observed for the energy uncertainties in Fig.~\ref{fig:si_uncertainty}c. At later iterations this decrease stagnates and the runs fluctuate around a low uncertainty. 

When comparing the uncertainty trends with the evolution of the energy errors relative to the GAP18 reference database in Fig.~\ref{fig:si_uncertainty}d, a consistent picture emerges.
The GAP18 database contains many configurations and phases of silicon that are thermodynamically irrelevant for the $(P,T)$ conditions explored in this work. Consequently, our AL strategy does not sample information about these regions, and accuracy with respect to them is naturally sacrificed.
Despite this, we observe a clear overall reduction in total energy error, indicating a global improvement of the MLIP throughout the AL iterations.
Minor irregularities in both the uncertainty estimates and the energy errors across iterations are plausibly attributable to variations in training performance, which we were not able to eliminate completely.

\begin{figure}
    \centering
    \includegraphics[width=.5\textwidth]{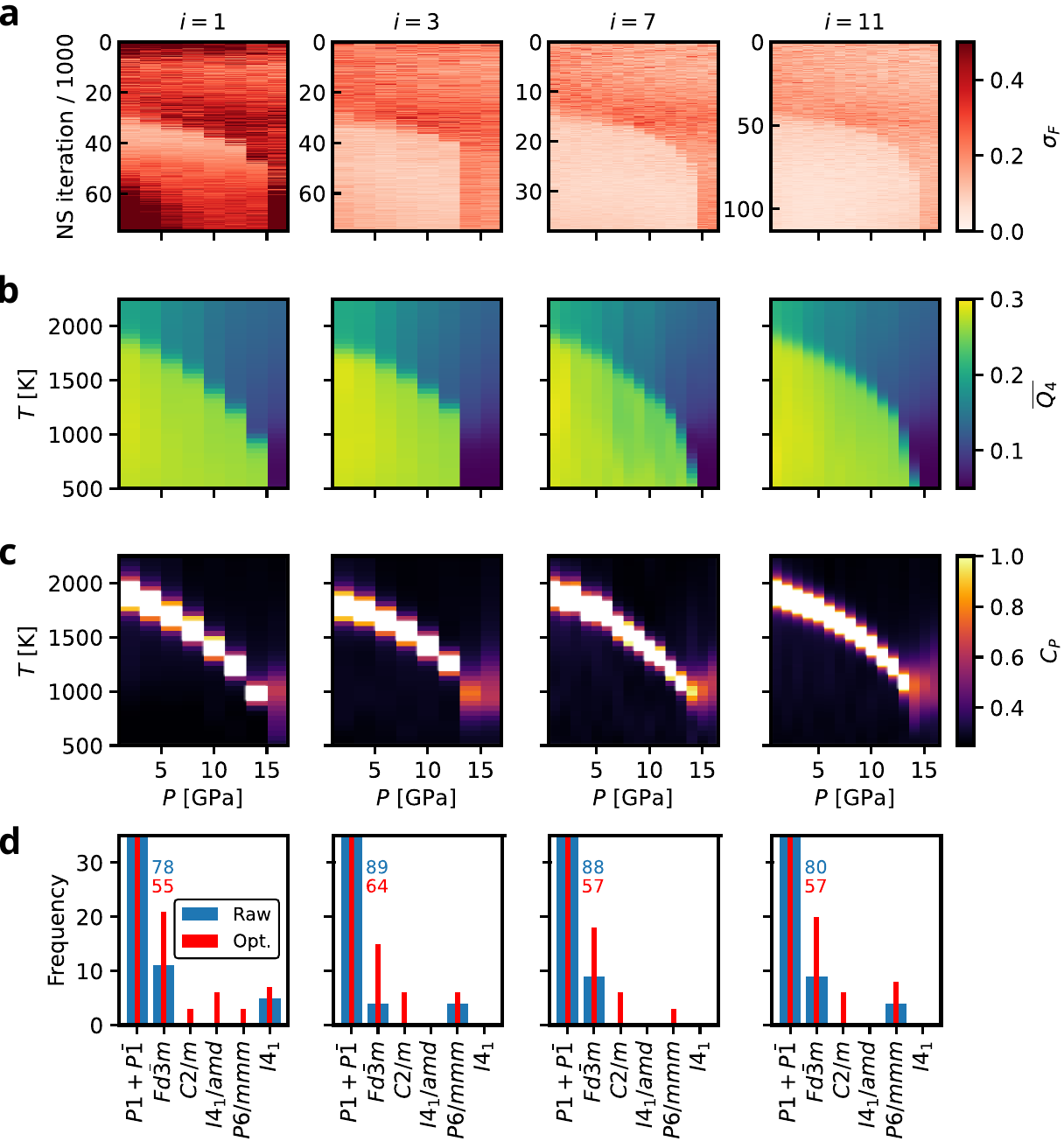}

    \caption{
        Monitoring of key quantities at different iterations $i$ of the AL strategy for silicon. 
        a) Normed force uncertainties of the NS sample trajectories (considering every 10th sample). 
        b) NS expectation values of the constant pressure heatcapacity. c) The mean of a Steinhardt bond order parameter $\overline{Q_4}$ order parameter. 
        d) Distribution of space groups of the $N_\mathrm{post}=100$ AL samples per iteration. Blue and red numbers indicate values for cut off bars.
        Normalized force uncertainty and $\overline{Q_4}$ are dimensionless and $C_P$ is given in units of $10^{-3} \, \mathrm{eV}\, \mathrm{K}^{-1} \, \mathrm{atom}^{-1}$.
    }
    \label{fig:si_dashboard}
\end{figure}
Fig.~\ref{fig:si_dashboard}a shows the normalized force uncertainties over the full $M$ sample trajectories. They reveal, that the model trained on the initial database alone ($i = 1$) exhibits large uncertainties, especially during the early and late stages of sampling and, notably, for the latter specifically at low and high pressures. From iteration \(i=3\) onward, we observe only a small, gradual reduction in uncertainty, consistent with the trend seen in the average force uncertainty, Fig.~\ref{fig:si_uncertainty}b. Moreover, across all iterations, the force uncertainties separate into two distinct magnitude regimes, which also explains the double-peaked structure visible in the force-uncertainty distribution in Fig.~\ref{fig:si_uncertainty}b.

For analyzing physical observables as a function of AL iteration, the mean of a Steinhardt bond order parameter over all atoms, $\overline{Q_4}$, is used as a structural order parameter distinguishing the different silicon phases, Fig.~\ref{fig:si_dashboard}b \cite{steinhardt_bond-orientational_1983,unglert_replica_2025}. Furthermore, we track the expectation values of the constant-pressure heat capacity as an indicator of first-order phase transitions, Fig.~\ref{fig:si_dashboard}c. 
Already after the first AL iteration (\(i=1\)), the expectation values exhibit a striking resemblance to the silicon phase diagram, revealing two solid phases and the characteristic negatively sloped melting line.  The low-pressure region is dominated by the cubic diamond $Fd\bar{3}m$ phase, while the simple hexagonal $P6/mmm$ phase constitutes the high-pressure ground state. At later AL iterations, a solid-solid phase transition from the simple hexagonal to the cubic diamond phase emerges at a pressure region around \SI{14}{GPa}. 

Additionally, we performed a coarse symmetry analysis of the $N_\mathrm{post}=100$ configurations sampled at each iteration using \texttt{spglib}~\cite{togo_textttspglib_2018} (see Fig.~\ref{fig:si_dashboard}d). 
Besides the raw samples, we also determined the symmetries of the corresponding optimized configurations, where the positional degrees of freedom were relaxed with the MLIP model of the current AL iteration. 
This analysis shows that the AL strategy samples a broad range of crystalline basins, while concentrating on those that are thermodynamically most relevant. 
Notably, even after positional optimization, the vast majority of configurations are assigned to the lowest-symmetry space groups $P1$ and $P\bar{1}$, indicating that the AL strategy predominantly samples higher-energy, more disordered structures. 
We attribute this to the construction of the initial database which contains no liquid configurations. 
While the most relevant crystalline basins are already reasonably well represented, we expect the AL strategy to collect additional information not only for the liquid phase, but also for the transition pathways connecting the thermodynamically dominant crystalline basins.
The symmetry analysis also reveals a shift in the sampled configurations for the high-pressure domain from the $\beta-Sn$ \(I4_1/amd\) phase towards the \(P6/mmm\) phase, which is the expected high-pressure ground state at the r2SCAN level of theory \cite{unglert_replica_2025}. This indicates that the initial model possessed an erroneous morphology that biased sampling toward the \(\beta\)-tin basin, and that this error was successfully corrected by the AL strategy after only a few iterations.
Figure~\ref{fig:si_initial_db} shows the atomic environments collected throughout the entire AL run. Beyond the clear tendency to sample liquid configurations, it also illustrates how regions associated with $\beta$-Sn and \(P6/mmm\) are successfully filled.

\subsection{Germanium}

\begin{figure}
    \centering
    \includegraphics[width=1.\columnwidth]{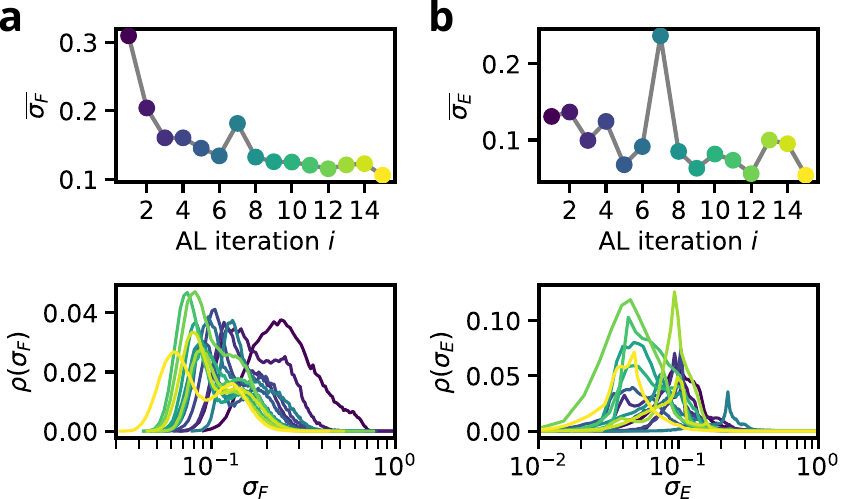}
    \caption{
        Distribution and averaged value of force and energy uncertainties for the Ge AL run.
        a) Force uncertainty over all NS sample trajectories. b) Energy uncertainty over all NS sample trajectories.
        Bottom panels show the  distribution of the respective quantity, top panels the average values for each AL iteration.
    }
    \label{fig:ge_uncertainty}
\end{figure}
The entries in the Materials Project database for germanium comprise phases similar to the silicon case (see Table~\ref{tab:initial_db}). 
Starting from this initial Ge database, the AL procedure was carried out using a similar RENS parameter schedule (see SI for details) and the same number of extracted structures per iteration, $N_\mathrm{post}=100$ as in the silicon case. 
Analyzing the two uncertainty metrics across the AL iterations in Fig.~\ref{fig:ge_uncertainty}, we observe convergence behaviour closely resembling that of silicon: only minor improvements in the overall uncertainty level occur after iteration $i=3$. 
While the average normalized force uncertainties in Fig.~\ref{fig:ge_uncertainty}a exhibit a well-controlled, nearly monotonic decrease, the average energy uncertainties in Fig.~\ref{fig:ge_uncertainty}b show somewhat less stability.

\begin{figure}
    \centering
    \includegraphics[width=1.\columnwidth]{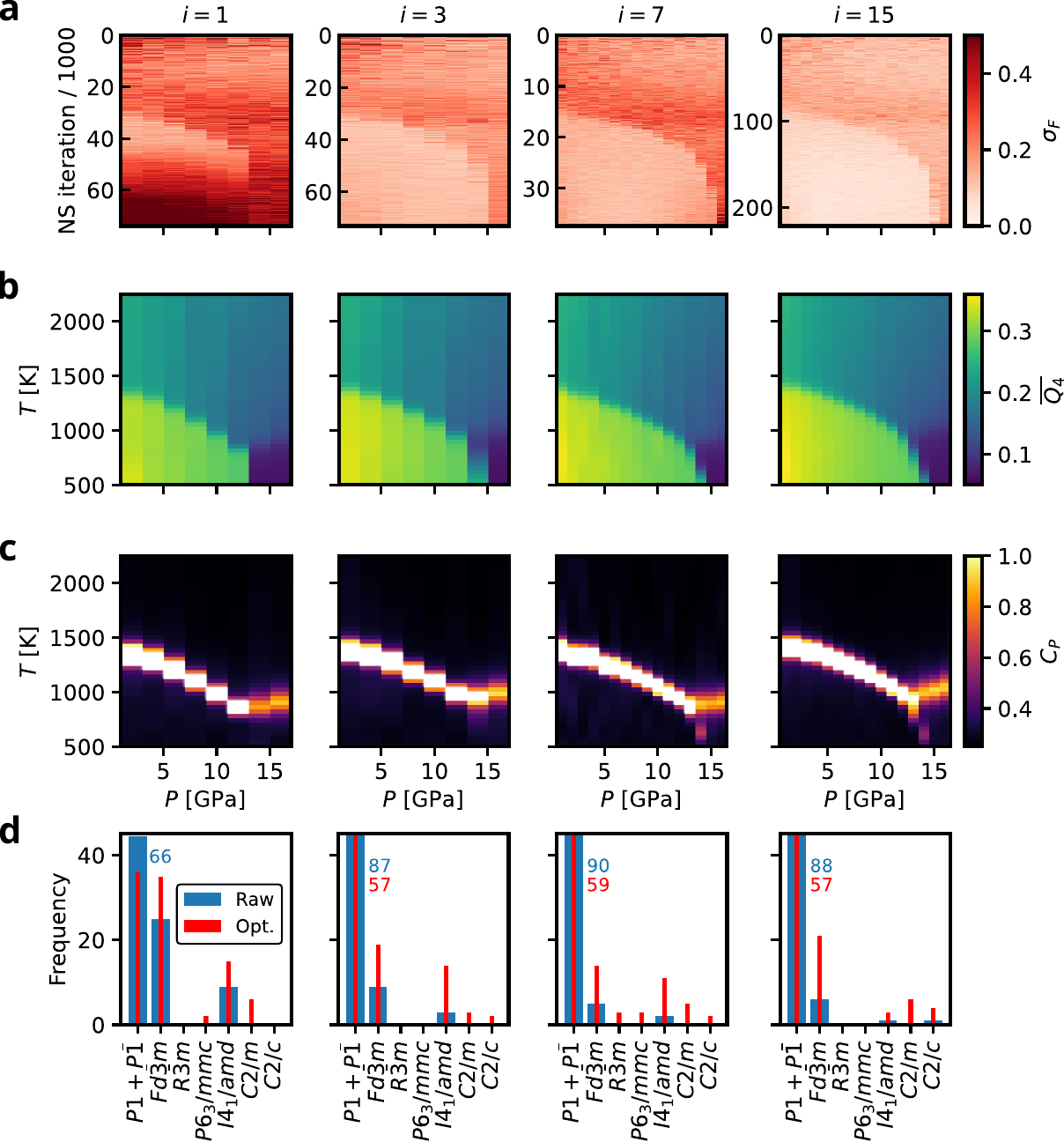}
    \caption{
        Summary of the germanium AL run for four exemplary iterations $i$ of the AL strategy for germanium. 
        a) Normed force uncertainties.
        b) NS expectation values of the constant pressure heat capacity $C_P$.
        c) The $\overline{Q_4}$ order parameter. 
        d) Distribution of space groups of the $N_\mathrm{post}=100$ AL samples per iteration. Blue and red numbers indicate values for cut off bars.
        Note, that the normalized force uncertainty and $\overline{Q_4}$ are dimensionless and $C_P$ is given in units of $10^{-3} \, \mathrm{eV}\, \mathrm{K}^{-1} \, \mathrm{atom}^{-1}$.
    }
    \label{fig:ge_dashboard}
\end{figure}
The normalized force uncertainty as a function of the NS iteration and the pressure replica in Fig.~\ref{fig:ge_dashboard}a shows particularly high values toward the final NS iteration at AL iteration $i=1$ across the entire pressure range. 
These deficiencies of the MLIP in the thermodynamically most relevant phases are quickly mitigated by the AL strategy. 
Consistent with Fig.~\ref{fig:ge_uncertainty}a, after iteration $i=3$ the force uncertainties decrease and subsequently exhibit only minor fluctuations.
This behaviour is also reflected in the symmetry analysis shown in Fig.~\ref{fig:ge_dashboard}d. 
After the first iteration, the sampled configurations predominantly belong to the two thermodynamically most relevant solid phases, $Fd\bar{3}m$ and $I4_1/amd$. 
In later iterations, however, the AL strategy increasingly samples disordered, higher-energy configurations—mirroring the trend already observed for silicon.

The computed thermodynamic observables in Fig.~\ref{fig:ge_dashboard}b and c show that the overall shape of the melting line— negatively sloped at low pressures and transitioning into positively sloped at higher—is captured already in the first iteration. 
From iteration $i=3$ onward, the emergence of the $Fd\bar{3}m \to I4_1/amd$ phase boundary becomes clearly visible. 
Subsequent iterations introduce only minor corrections to the precise location of these phase boundaries.

\begin{figure}
    \centering
    \includegraphics[width=1\columnwidth]{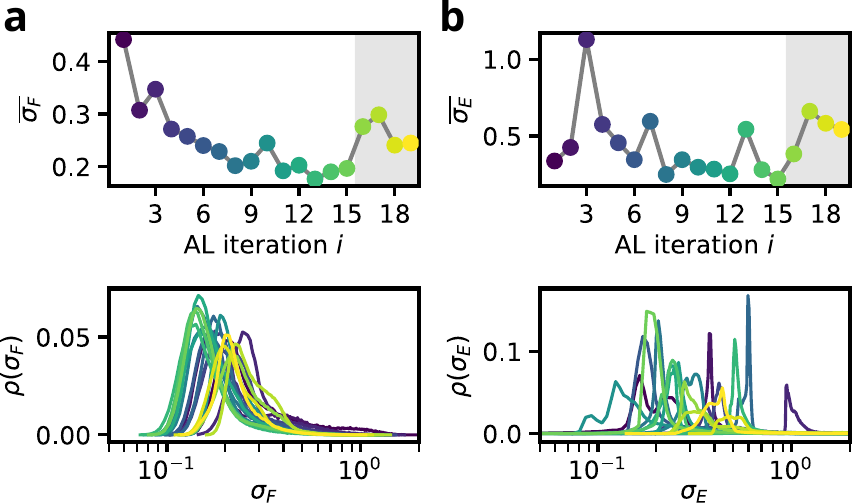}
    \caption{
    Distribution and averaged value of force and energy uncertainties for the Ti AL run.
    a) Force uncertainty over all NS sample trajectories. b) Energy uncertainty over all NS sample trajectories.
    Bottom panels show the  distribution of the respective quantity, top panels the average values for each AL iteration.
    The shaded area corresponds to the iterations that were conducted with an increased system size of $N_\mathrm{atoms} = 32$
    }
    \label{fig:ti_uncertainty}
\end{figure}

\begin{figure}
    \centering
    \includegraphics[width=1.\columnwidth]{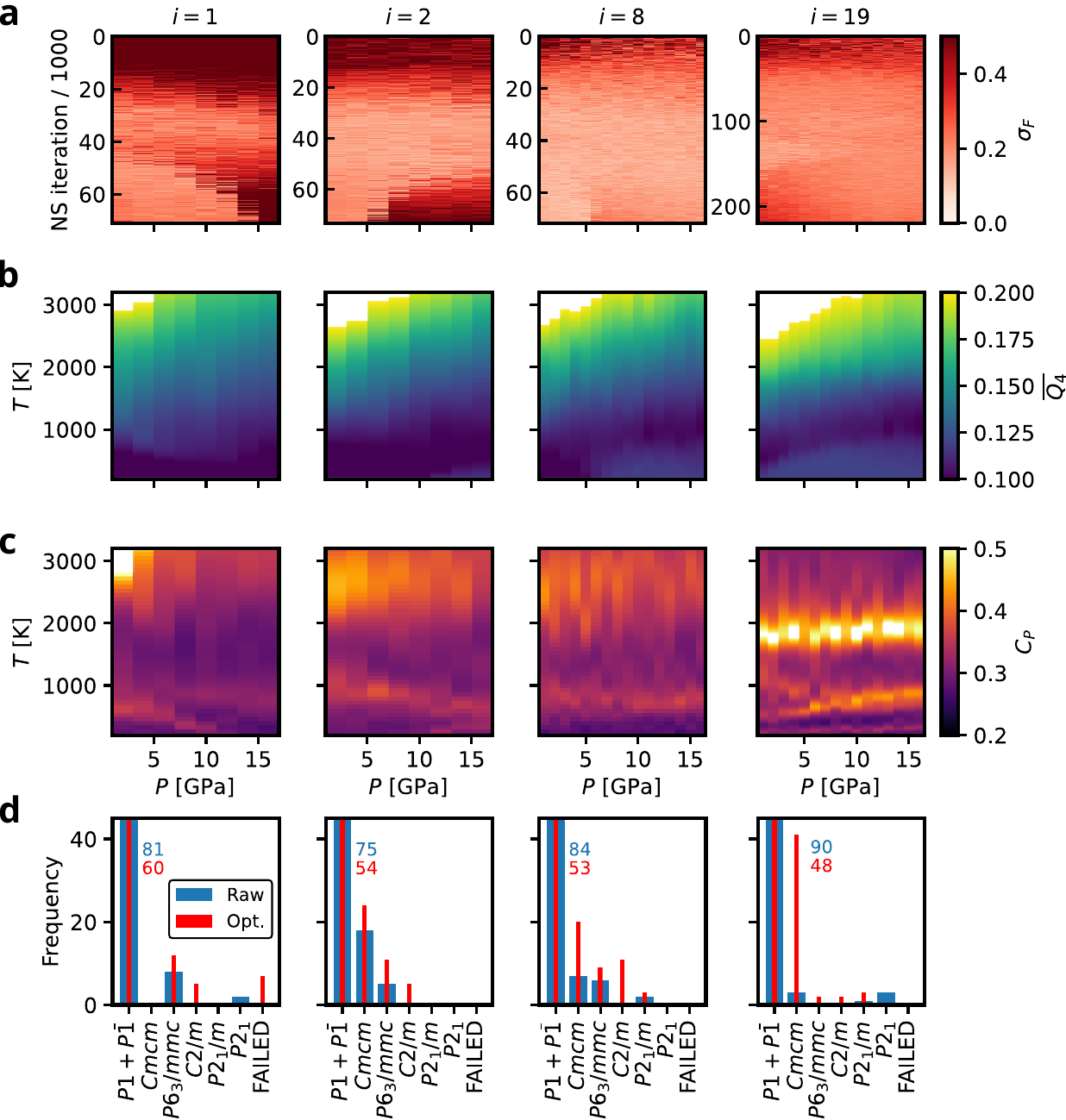}
    \caption{
    Summary of the Ti AL run. 
    a) Normed force uncertainties at different iterations $i$ of the AL strategy. 
    b) and c) NS expectation values of the constant pressure heatcapacity and the $\overline{Q_4}$ order parameter, respectively. 
    d) Distribution of space groups of the $N_\mathrm{post}=100$ AL samples per iteration. Blue and red numbers indicate values for cut off bars.
    Note, that the normalized force uncertainty and $\overline{Q_4}$ are dimensionless and $C_P$ is given in units of $10^{-3} \, \mathrm{eV}\, \mathrm{K}^{-1} \, \mathrm{atom}^{-1}$. In the first iteration for seven  sampled configurations either the geometry optimization failed numerically or the symmetry analysis broke down due to atoms approaching closer than \SI{0.5}{\angstrom}.
    }
    \label{fig:ti_dashboard}
\end{figure}

\subsection{Titanium}

For titanium the initial database contains only structures based on three distinct space groups ($P6_3/mmc$, $Im\bar{3}m$ and $P6/mmm$),  which represent the experimentally observed $\alpha$, $\beta$, and $\omega$ phases in this pressure regime  (see Table~\ref{tab:initial_db}). Starting from this initial database, the AL procedure was carried out with $N_\mathrm{post}=100$ extracted structures per iteration. 
Analyzing the two uncertainty metrics across the AL iterations in Fig.~\ref{fig:ti_uncertainty}, we observe convergence behaviour similar to that of silicon and germanium. 
In line with the results for Si and Ge and with Ref.~\cite{heid_spatially_2024}, the energy uncertainties for Ti, Fig.~\ref{fig:ti_uncertainty}b, are comparatively less reliable as proxies for the true error than the corresponding force uncertainties, Fig.~\ref{fig:ti_uncertainty}a.

Titanium differs from the previously discussed group~IV elements in that it behaves as a classical metal throughout the entire investigated pressure range between 1 and \SI{16}{GPa}. 
Due to this metallic nature, the phase transitions are noticeable less sharp than in silicon and germanium, Fig.~\ref{fig:ti_dashboard}b and c. 
While this complicates a detailed interpretation of the computed observables, it simultaneously facilitates a more straightforward scaling to larger system sizes. 
In particular, the broader transitions allow RENS simulations with $N_\mathrm{atoms} = 32$ to be performed without increasing the number of replicas $M$, thereby increasing the computational cost only by the expected factor of four. For a more detailed discussion on the scaling with system size in RENS as well as the exact parameter schedule we refer to the SI.

We therefore continued the AL procedure for four additional iterations (see the grey-shaded regions in Fig.~\ref{fig:ti_uncertainty}) using $N_\mathrm{atoms} = 32$. 
As anticipated, the AL MLIP exhibits a reasonable degree of transferability: the average force uncertainty in  Fig.~\ref{fig:ti_uncertainty}a does not diverge with increasing system size but instead stabilizes at a level comparable to that reached after five AL iterations for the smaller system. 
This behavior is consistent with Eq.~\eqref{eq:force_uncty}, which defines the force uncertainty as an intensive quantity. 

The normalized force uncertainties along the NS trajectories (Fig.~\ref{fig:ti_dashboard}a) show significantly elevated values during the first AL iterations. 
We attribute this to the fact that the initial database contains substantially less structural variety than in the silicon and germanium cases, rendering the learning of high-energy, disordered environments more challenging. 
These deficiencies of the initial model are also visible in the symmetry analysis for $i = 1$ in Fig.~\ref{fig:ti_dashboard}d, which shows that for seven of the $N_\mathrm{post} = 100$ sampled configurations the geometry optimisation failed numerically or the symmetry analysis broke down due to atoms approaching closer than \SI{0.5}{\angstrom}. 
This indicates that the initial potential contains significant artifacts, which drive some optimisations into unphysical regions. 
Nevertheless, since no such failures occur in subsequent iterations, these deficiencies appear to be sealed effectively after only a single AL cycle.

Figures~\ref{fig:ti_dashboard}b and c display the expectation values of $\overline{Q_4}$ and $C_P$, respectively. 
The increased system size in the last column results in a pronounced sharpening of the transitions, which is reflected in both observables. 
While the heat capacity exhibits more distinct features already indicative of the coexistence lines in the phase diagram, the $\overline{Q_4}$ expectation value clearly signals the presence of two solid phases—a low-temperature and a high-temperature phase—spanning the entire pressure range. 
Furthermore, a significant shift of the melting line toward lower pressures can be observed, a commonly encountered finite-size effect in NS simulations~\cite{unglert_neural-network_2023, Marchant_NPJCM23}.

\subsection{Phase diagrams}

Using the AL MLIPs described above, we now present accurate RENS simulations of the Si, Ge, and Ti phase diagrams at the r2SCAN level of theory, and compare them to reported results from the literature.

\begin{figure*}
    \centering
    \includegraphics[width=1.\textwidth]{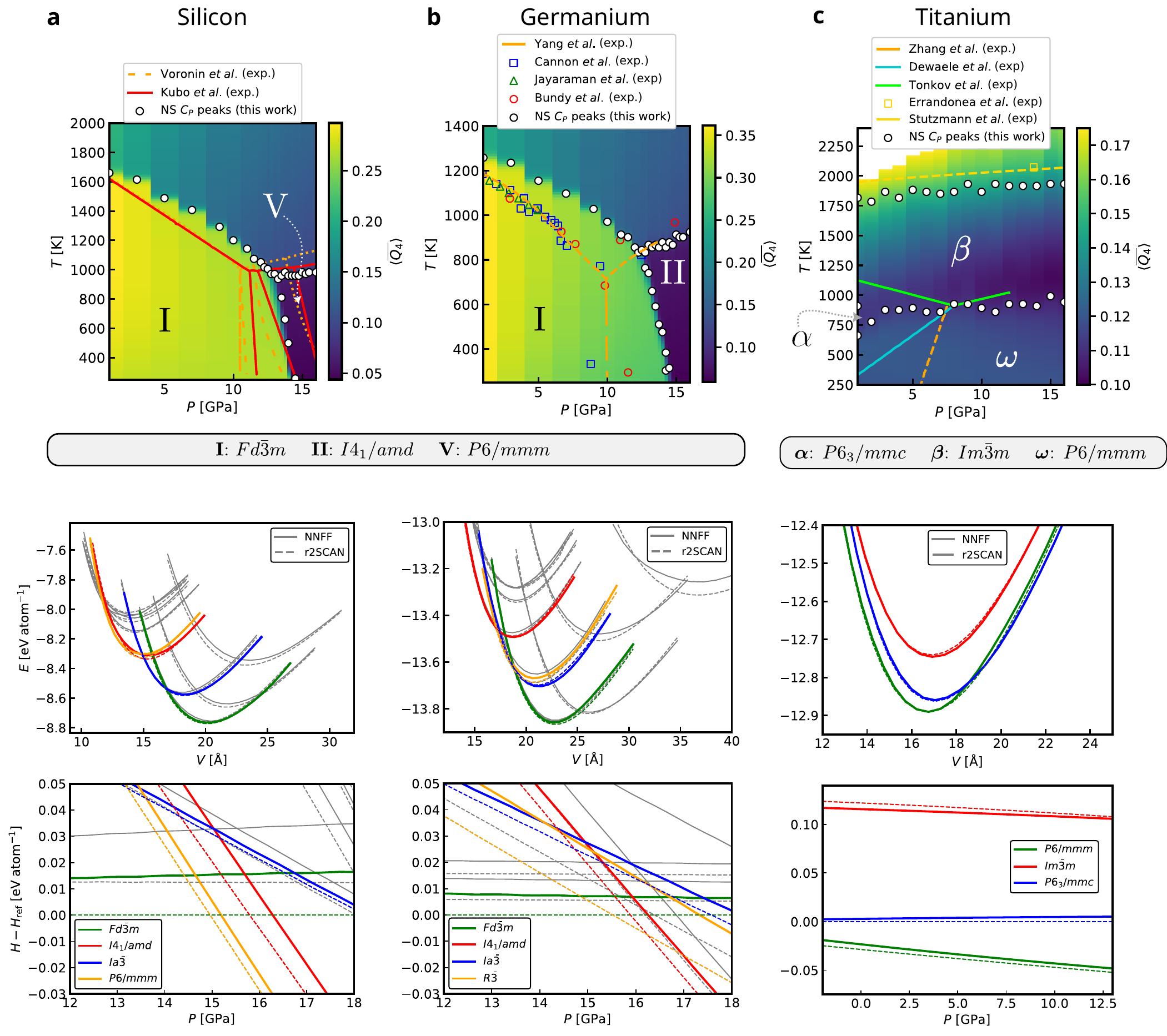}
    \caption{
    Proposed phase diagrams for the three investigated elemental systems. 
    a) Silicon with phase boundaries proposed from experiments according to Voronin \textit{et al.} \cite{voronin_situ_2003} and Kubo \textit{et al.} \cite{kubo_melting_2008}. 
    b) Germanium  with experimentally determined transition points from Cannon \textit{et al.} \cite{cannon_behavior_1974}, Jayaraman \textit{et al.} \cite{jayaraman_melting_1963}, Bundy \textit{et al.} \cite{bundy_phase_2004} and Yang \textit{et al.} \cite{yang_temperaturepressure_2004}. 
    c) Titanium. Experimental data for the melting line were taken from Errandonea \textit{et~al.}~\cite{errandonea_systematics_2001}, with a fit from Stutzmann \textit{et~al.}~\cite{stutzmann_high-pressure_2015}, while experimental data for the solid–solid transitions are taken from Tonkov \textit{et~al.}~\cite{tonkov_phase_2004} ($\beta \rightarrow \alpha$ and $\beta \rightarrow \omega$) and from Zhang \textit{et~al.}~\cite{zhang_experimental_2008} and Dewaele \textit{et~al.}~\cite{dewaele_high_2015} ($\alpha \rightarrow \omega$). 
    Upper row panels show expectation value of structural order parameter $\overline{Q_4}$. Black circles show transition temperatures determined from $C_P$, in case of titanium from a gaussian mixture model fit. 
    Middle row panels show energy-volume curves evaluated with DFT and the model obtained from the corresponding AL strategy used to compute the phase diagrams. Enthalpy curves in the bottom row panels are obtained by fitting a Birch-Murnaghan equation of state to the energy-volume curves.
    }
    \label{fig:final_phase_diagrams}
\end{figure*}

For silicon, the sharp transitions resulted in a clear picture of the phase diagram already for the relatively small system with $N_\mathrm{atoms}=16$, Fig.~\ref{fig:si_dashboard}b and c.
Nevertheless, some finite-size effects like the overestimation of the melting temperature remain, hence we performed an additional simulation  with $N_\mathrm{atoms}=32$ and $K=800$ and $M=24$, decreasing the replica pressure intervals around the transition region. 

The resulting phase diagram is shown in Fig.~\ref{fig:final_phase_diagrams}a together with the experimental results.
The solid phase in the pressure range up to approximately \SI{15}{GPa} corresponds to the $Fd\bar{3}m$ structure, which is separated by a negatively sloped phase boundary from the high-pressure $P6/mmm$ phase. 
Both phases extend down to \SI{0}{K} and form the respective ground states. 
The RENS simulation captures this behaviour, and the corresponding NS sample trajectories terminate in the perfectly crystalline $Fd\bar{3}m$ and $P6/mmm$ structures. 
As before, we use the expectation value of the $\overline{Q_4}$ order parameter, which effectively distinguishes the liquid and both solid phases. 

In the middle panel of Fig.~\ref{fig:final_phase_diagrams}a, we additionally show energy–volume curves for all silicon phases contained in the initial database, evaluated both with DFT (dashed lines) and the AL MLIP (solid lines). 
By fitting an equation of state to these curves, we obtain the ground-state enthalpies of each phase, which allow us to estimate phase stability at \SI{0}{K}. 
The results demonstrate that the AL model retains the ability to describe the ordered crystalline phases even after the AL procedure, which primarily enriches the dataset with finite-temperature configurations exhibiting thermal distortions of both atomic coordinates and cell parameters. 

\begin{figure*}
    \centering
    \includegraphics[width=1.\textwidth]{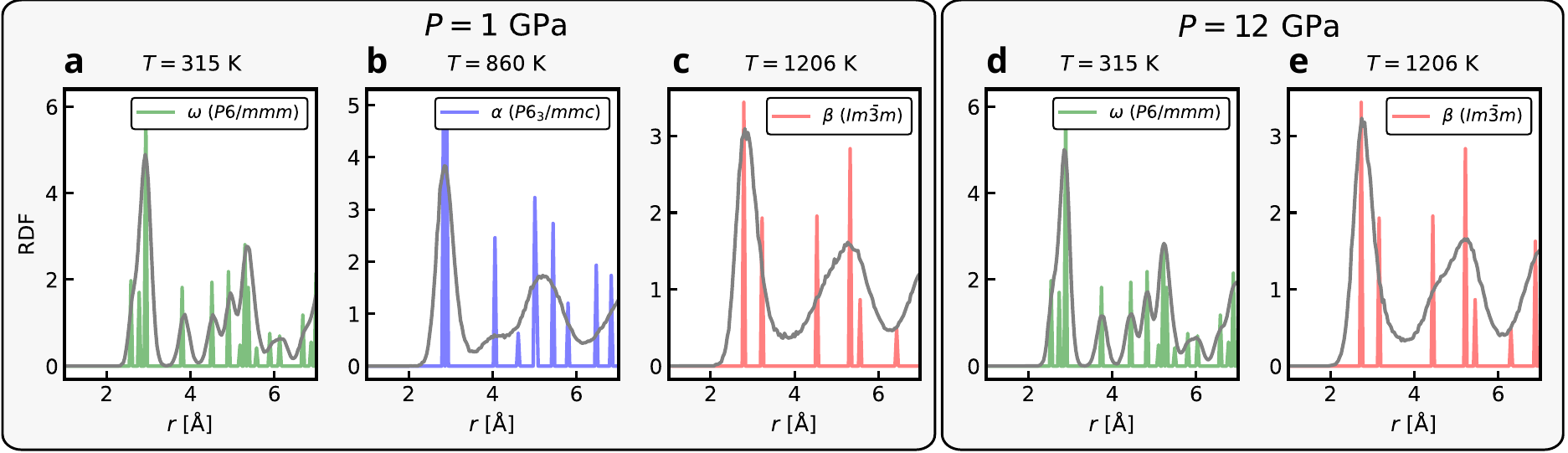}
    \caption{
    Identification of titanium phases by NS expectation values of the radial distribution function (RDF) for \SI{1}{GPa} and \SI{12}{GPa} at selected temperatures.
    }
    \label{fig:titanium_phases}
\end{figure*}

Comparison of the resulting silicon phase diagram with literature data~\cite{voronin_situ_2003,kubo_melting_2008,paul_anharmonic_2019} shows good agreement for the melting line. 
The melting temperature for the $liq. \rightarrow Fd\bar{3}m$ transition is slightly overestimated while the $liq. \rightarrow P6/mmm$ transition is slightly underestimated. 
The latter transition exhibits a much smaller latent heat and is therefore, according to our experience, less affected by finite-size errors. 
In the solid region, literature reports a sequence of transitions $Fd\bar{3}m \rightarrow I4_1/amd \rightarrow R\bar{3} \rightarrow P6/mmm$ between 10 and \SI{17}{GPa}. 
This differs from our r2SCAN-based prediction, which shows only a direct $Fd\bar{3}m \rightarrow P6/mmm$ transition. 
We attribute this discrepancy to limitations of r2SCAN in capturing the finite-temperature properties of the intermediate phases. 
As evident from the bottom panel of Fig.~\ref{fig:final_phase_diagrams}a, both the DFT and MLIP energy–volume data predict only $Fd\bar{3}m$ and $P6/mmm$ as stable ground-state phases, with a transition near \SI{15}{GPa}, consistent with our RENS results. 
Earlier findings using the PBE functional differ from the present mainly in terms of the presence of several intermediate phases above 10~GPa \cite{paul_anharmonic_2019,unglert_neural-network_2023} demonstrating substantial differences between the PBE and r2SCAN energy landscapes for silicon.

Figure~\ref{fig:final_phase_diagrams}b shows the final phase diagram for germanium, obtained from a RENS simulation with $N_\mathrm{atoms}=32$, $K=800$, and $M=24$, compared to experimentally determined phase boundaries from Refs.~\cite{yang_temperaturepressure_2004,cannon_behavior_1974,jayaraman_melting_1963,bundy_phase_2004}. 
The phase diagram consists of a liquid phase and two solid phases: $Fd\bar{3}m$ at lower pressures and $I4_1/amd$ at higher pressures. 
This is qualitatively consistent with experimental observations, although the solid–solid transition appears shifted towards slightly higher pressures. 
The energy–volume and enthalpy data (middle and bottom panels of Fig.~\ref{fig:final_phase_diagrams}b) clarify this behaviour: 
r2SCAN predicts the $Fd\bar{3}m \rightarrow I4_1/amd$ transition near \SI{16}{GPa} at \SI{0}{K}. 
Despite these deviations in absolute position, our results support the experimentally observed negatively sloped boundary \cite{bundy_phase_2004}, and contrast the earlier reported positively sloped phase boundary \cite{cannon_behavior_1974}.

The titanium phase diagram marks a departure from the group~IV elements, as titanium exhibits metallic bonding throughout the entire investigated pressure range. 
The phase diagram, obtained from a RENS simulation with $N_\mathrm{atoms}=32$, $K=1200$, and $M=16$ equally spaced replicas in the range from 1 to \SI{16}{GPa}, is shown in Fig.~\ref{fig:final_phase_diagrams}c. 

Our computed phase diagram reproduces all four relevant phases: a liquid phase separated by a positively sloped melting line from the high-$T$ $\beta$ phase, the low-$T$ $\omega$ phase, and the intermediate $\alpha$ phase, which can be regarded as a structural bridge between the former two. 
In agreement with experimental results \cite{tonkov_phase_2004,zhang_experimental_2008,dewaele_high_2015}, the $\alpha$ phase forms only a narrow stability region around \SI{800}{K} enclosed by the $\beta$ and $\omega$ phases and ending in a triple point. Our simulation predicts the triple point at around \SI{2}{GPa} at a significantly lower pressure than the experimental value around 7~GPa. This could indicate a small spurious r2SCAN preference for the low volume $\omega$-phase.
Earlier DFT calculations based on the PBE functional predict a triple point at around 11~GPa \cite{mei_density-functional_2009}, which could reflect this functionals preference for large volume phases. Furthermore, it should be pointed out that because all solid phases of titanium are metallic and the energetic differences are comparable small, the electronic free energy plays a significant role in determining the relative stability of the $\alpha$ and $\omega$ phases. According to Ref.~\cite{mei_density-functional_2009}, the electronic entropy accounts for up to 30\% of the total entropy change for the $\alpha \rightarrow \beta$ transition. 
Including such electronic effects within NS remains an open challenge.

Interestingly, the $C_P$ expectation values recorded during the AL strategy (see Fig.~\ref{fig:ti_dashboard}c), particularly at iteration $i=19$, show a similar topology but with the $\alpha$ region significantly more extended at lower pressures. While this appears to align better with the experiment, we found that this behaviour originates from slight imperfections in the MLIP fit, which systematically underpredicted the enthalpy differences between the $\alpha$ and $\omega$ phases in the simulations shown in Fig.~\ref{fig:ti_dashboard}c, thereby either artificially stabilizing the $\alpha$ phase or destabilizing the $\omega$ phase. 
The $\alpha$-region phase boundaries are therefore highly sensitive to small energetic deviations. The phase diagram for Ti shown in Fig.~\ref{fig:final_phase_diagrams}c was computed using the final force field, which also incorporates information sampled during iteration $i=19$, and shows excellent agreement with remaining deviations in the enthalpy on the order of only a few \SI{}{meV/\mathrm{atom}}. We therefore conclude that our results represent an accurate prediction on the level of r2SCAN of the phase diagram of titanium neglecting electronic degrees of freedom. 

The phase transitions in titanium are less sharp than those in the group~IV elements. Furthermore, similar densities and structural resemblance of the phases make it more difficult to distinguish them using a simple scalar order parameter like $\overline{Q_4}$.
To confirm the phase identities, we computed expectation values of the radial distribution function (RDF) at selected temperatures for $P = 1$ and \SI{12}{GPa}, shown in Fig.~\ref{fig:titanium_phases}. 
The resulting RDFs are compared to reference RDFs of the perfectly crystalline $\alpha$, $\beta$, and $\omega$ phases. 
At both pressures, the lower-symmetry $\omega$ phase is clearly identifiable as the ground state at the low temperature of \SI{315}{K} (see Fig.~\ref{fig:titanium_phases}a and d). 
At the higher temperature of \SI{1206}{K}, the $\beta$ phase can be recognized for both pressures (see Fig.~\ref{fig:titanium_phases}c and e). 
Despite substantial thermal broadening, two characteristic features remain visible: a left main peak around \SI{3}{\angstrom} with a slight shoulder on its right side, and a second peak around \SI{5}{\angstrom} exhibiting a corresponding shoulder on the left.
At \SI{1}{GPa}, one also observes signatures of the $\alpha$ phase at the intermediate temperature of \SI{860}{K} (see Fig.~\ref{fig:titanium_phases}b). 
Although the RDF is significantly smeared by thermal fluctuations, a distinct peak around \SI{4}{\angstrom} unambiguously identifies the $\alpha$ phase. 
Inspection of the remaining two dominant peaks further distinguishes it from both $\beta$ and $\omega$.

\section{Discussion}

In this work, we demonstrated that a RENS-based active learning (AL) strategy enables the automated computation of entire $P$--$T$ phase diagrams, making it an attractive framework for autonomous materials discovery. While this work focused on the r2SCAN exchange--correlation functional for the underlying DFT evaluations, the proposed framework is readily extendable to other functionals, and we advocate their exploration in future applications of our AL approach to phase-diagram prediction.



An open question concerns the amount of \emph{a priori} information that must be provided to the initial model to ensure exploration of all relevant regions of the PES during iterative AL. 
Recent advances in state-of-the-art $\mathrm{E}(3)$-equivariant graph neural networks alleviate some of the data hunger of first-generation MLIPs 
and may render RENS-based AL strategies viable even for materials where initial data are scarce. 
Moreover, the emergence of foundation models in atomistic simulation provides architectures with strong generalization capabilities, in which structural and chemical regularities are already encoded from vast datasets. 
In this context, RENS-based AL strategies offer a principled and unbiased framework for the fine-tuning of such pretrained models to specific materials~\cite{radova_fine-tuning_2025}.

Nevertheless, the NS computational cost remains a major bottleneck, particularly when coupled with expensive $\mathrm{E}(3)$-equivariant models. 
Continued improvements in NS algorithms, faster model evaluations, and advances in computing hardware are  expected to play a crucial role in bringing RENS-based AL strategies for complex materials systems within practical reach.

\section{Methods}

\subsection{Replica-exchange nested sampling}

Nested sampling evolves a population of $K$ walker configurations through configuration space by iteratively removing hypershells of decreasing configuration-space volume. 
In each iteration, the configuration with the lowest likelihood is removed and recorded as a sample. 
Its likelihood defines a threshold, which in turn specifies a likelihood-constrained distribution from which a new configuration must be drawn independently. 
This construction ensures that the remaining configuration-space volume contracts according to a beta distribution, leading to an exponential decrease in accessible volume. 
Each sample $R_i$ can thus be associated with the hypershell removed at iteration $i$ (see Fig.~\ref{fig:al_strategies}a). 
Because each sample is assigned an associated configuration-space volume, NS enables the direct and efficient evaluation of configuration-space integrals and, consequently, the computation of thermodynamic expectation values. 
Throughout this work we will refer to these configurations as NS samples.
By progressing in exponential steps of configuration-space volume, NS naturally traverses phase transitions uniformly, making it particularly powerful for studying first-order transitions that involve abrupt and large-scale changes in accessible configuration space.

By construction of the algorithm, NS is athermal and hence the acquired samples can be used to compute thermodynamic observables at any temperature.
Consequently, only a few well-chosen NS simulations at a range of pressures are sufficient to map out a material’s $p$–$T$ phase diagram.

In a recent study~\cite{unglert_replica_2025}, we demonstrated that the efficiency of the likelihood-constrained sampling can be substantially improved by introducing a replica-exchange (RE) mechanism that couples several NS simulations performed at different external conditions. 
A replica-exchange nested sampling (RENS) simulation consists of $M$ replicas, each carrying out an independent NS run but at a different external pressure $P_m$, $m = 1,\dots,M$. 
All replicas evolve simultaneously and independently according to the standard NS steps, but they are additionally coupled at regular intervals through swap moves reminiscent of parallel tempering. 
Whereas parallel tempering couples the sampling processes of several canonical distributions, RENS instead couples the distinct likelihood-constrained distributions—i.e.\ the constrained priors that define the distributions of the $K$ walkers within each replica—that arise from the different external pressures. 
This coupling enables a substantial increase in ergodicity by allowing configurations to move between replicas and thus explore neighbouring regions of the enthalpy landscape.

The frequency at which such swap moves succeed—the swap acceptance rate—is an important indicator of how efficiently neighbouring replicas communicate. 
High acceptance rates signal strong overlap between the likelihood-constrained distributions of adjacent pressures, whereas low acceptance rates indicate insufficient overlap and limited mixing. 
Hence, RENS requires a careful choice of the pressure intervals to prevent a loss of overlap. 
Since RENS incurs only negligible computational overhead, we employ it exclusively throughout this work.
Note that the RE mechanism in RENS should be regarded solely as an auxiliary enhancement of the MCMC-based likelihood-constrained sampling. 
The thermodynamic interpretation of each replica in a RENS simulation remains identical to that of an independent NS run. 
Accordingly, we do not distinguish strictly between RENS and NS unless explicitly required.

The RENS simulations presented in this work were performed using our custom \texttt{JAXNEST} package, a Python code specifically designed for JAX-based MLIPs. 
To cope with the computational cost of the RENS simulations, we implemented a multi-GPU parallelization scheme that distributes the workload across several devices.

In the multi-GPU implementation, the replica-exchange logic was slightly adapted to minimize inter-device communication overhead. 
For a simulation comprising $M$ replicas executed on $n_\mathrm{GPU}$ GPUs, we assign $M/n_\mathrm{GPU}$ replicas to each GPU. 
Two types of RE moves are employed, denoted as \emph{intra}- and \emph{inter}-swap moves.
\emph{Intra}-swap moves exchange configurations between replicas residing on the same GPU and are performed during the MCMC random walks of each NS iteration, following the procedure described in Ref.~\cite{unglert_replica_2025}. 
\emph{Inter}-swap moves, in contrast, are executed between NS iterations and only every $I_\mathrm{inter}$-th iteration. 
In these moves, all replicas are gathered on the root GPU, where global swaps between all replicas are attempted, again following the scheme outlined in Ref.~\cite{unglert_replica_2025}.

A comprehensive description of the principal parameters controlling RENS simulations is provided in Ref.~\cite{unglert_replica_2025}. 
For completeness, the specific parameter values used in this work are summarized in Table~\ref{tab:rens_params}.

\begin{table*}[t]
    \centering
    \setlength{\tabcolsep}{3pt}
    \begin{tabular}{l c c c c c c c c c c c } 
        \hline\hline
            prior & 
            walker init. & 
            $V_\mathrm{min}$ & 
            $V_\mathrm{max}$ & 
            $P_\mathrm{acc}$ window &
            $N_\mathrm{adjust}$ &
            $f_\mathrm{adjust}$ &
            step types          &
            step ratio          &
            $d_0$               &
            $N_\mathrm{atoms}$ 
            \\
            \hline
            volume &
            triclinic grid &
            \SI{10.}{\angstrom^3 \; \mathrm{atom}^{-1}} &
            \SI{52.7}{\angstrom^3 \; \mathrm{atom}^{-1}} &
            (0.25, 0.75) &
            400 &
            1.5 &
            \makecell{GMC, AP-MC\\ volume, stretch, shear} &
            1:8:16:8:8 &
            0.9 &
            16
    \end{tabular}
    \caption{Employed default NS parameters employed throughout this work if not specified else. A detailed description of the parameters can be found in Ref.~\cite{unglert_replica_2025}.}
    \label{tab:rens_params}
\end{table*}


\subsection{Initial database creation}

To construct the initial databases, we extracted all $n_\mathrm{MP}$ experimentally reported entries of the target material from the Materials Project \cite{jain_commentary_2013, horton_accelerated_2025}. To enrich the structural diversity, each configuration was modified in three ways: (i) A supercell was created that fits a predefined cutoff $R^\mathrm{init}_\mathrm{cut}$ (ii) isotropic volume scaling using a predefined set of $n_\mathrm{scaling}$ scaling factors $\{ f^i \}_{i=1}^{n_\mathrm{scaling}}$ were performed, and (iii) atomic displacements were generated by adding independent Gaussian noise with $n_\mathrm{\sigma}$ specified standard deviations $\{ \sigma^i_\mathrm{rattle} \}_{i=1}^{n_\mathrm{\sigma}}$. 
The total number of configurations created this way for a given material adds up to $n_\mathrm{total}=n_\mathrm{MP} \cdot n_\mathrm{scaling} \cdot n_\sigma$.
All  $n_\mathrm{total}$ configurations were labeled using the DFT parametrization specified in the following.
Table \ref{tab:initial_db} provides an overview of the configuration types and the chosen diversification parameters.

\begin{table*}[t]
    \centering
    \setlength{\tabcolsep}{6pt}
    \renewcommand{\arraystretch}{1.2}
    \begin{tabular}{l p{4.5cm} p{4.5cm} p{4.5cm}} 
        \hline\hline
          & Si
          & Ge
          & Ti
        \\
        \hline
        $R^\mathrm{init}_\mathrm{cut} [\si{\angstrom}]$ &
            3.0 & 
            3.0 & 
            3.0 
        \\
        $\{ f^i \}_{i=1}^{n_\mathrm{scaling}}$ &
            $\mathrm{linspace}(0.9, 1.1, 20)$ & 
            $\mathrm{linspace}(0.9, 1.1, 20)$ & 
            $\mathrm{linspace}(0.9, 1.1, 20)$ 
        \\
        $\{ \sigma^i_\mathrm{rattle} \}_{i=1}^{n_\mathrm{\sigma}}$ $[\si{\angstrom}]$&
            \{0.0, 0.2\} &
            \{0.0, 0.2\} &
            \{0.0, 0.05, 0.1, 0.2\} 
        \\
        $n_\mathrm{MP}$ &
            12 & 14 & 3
        \\
        $n_\mathrm{total}$ &
             480 & 560 & 240 
        \\
        \hline
        spacegroups MP &
            $I4/mmm$, $I4_1/amd$, $R\overline{3}$, $P6/mmm$, $P6_3/mmc$, $P6_3/mmc$, $Ia\overline{3}$, $Fm\overline{3}m$, $Fd\overline{3}m$, $Fd\overline{3}m$, $Cmcm$, $Cmce$ &
            $I4_1/amd$, $R\overline{3}$, $P6_3mc$, $P6_3/mmc$, $P6_3/mmc$, $P6_3/mmc$, $Ia\overline{3}$, $Fm\overline{3}m$, $Fd\overline{3}m$, $Fd\overline{3}m$, $Fd\overline{3}m$, $Cmce$, $Imma$, $P4_32_12$ &
            $P6_3/mmc$, $P6/mmm$, $Im\bar{3}m$
    \end{tabular}
    \caption{Details on diversification parameters for initial database  creation as well as spacegroups contained in seed structures from Materials Project. Note, that different phases from Materials Project entries may share the same spacegroups.}
    \label{tab:initial_db}
\end{table*}

\subsection{Density functional theory}

For the DFT calculations the r2SCAN \cite{furness_accurate_2020} functional as implemented in VASP \cite{kresse_efficiency_1996,kresse_ultrasoft_1999} was used. 
The cutoff energy for the plane wave basis was chosen as \SI{300}{eV}.
The partial occupancies for the orbitals were determined employing Fermi smearing with a smearing parameter of \SI{0.025}{eV}.
The reciprocal space sampling was performed on a Monkhorst-Pack grid with a \textit{k}-spacing of \SI{0.2}{\angstrom^{-1}} and the energy convergence criterion was set to \SI{e-7}{eV}. 

\subsection{Machine-learning interatomic potential} \label{sec:nnff}

The simulations in this work use our \texttt{NeuralIL} architecture \cite{montes-campos_differentiable_2022,carrete_deep_2023}.
Atomic coordinates are encoded into atom-centered descriptors that are invariant with respect to global rotations and translations \cite{bartok_representing_2013}, with relative positions of neighbors transformed into second-generation spherical Bessel descriptors \cite{kocer_continuous_2020}.
The descriptors are fed into a dense $128\times64\times32$ ResNet-inspired feature extractor \cite{he_deep_2016,carrete_deep_2023} using a Swish-1 differentiable activation function \cite{ramachandran_searching_2017}.
The implementation uses \texttt{JAX} \cite{bradbury_jax_2018} for just-in-time compilation and automatic differentiation, and \texttt{FLAX} \cite{heek_flax_2020} for simplified model construction and parameter bookkeeping.
For the descriptor computation we use a cutoff radius $r_\mathrm{c}=4.5$ to determine neighboring atoms and $n_\mathrm{max}=6$, controlling the number of basis function onto which the neighbor densities are projected.

We use \textsc{VeLO} \cite{metz_velo_2022} for the training of our committee models, which enables fast parameter optimization during AL cycles within only 100 epochs. 
Since RENS is exploring the entire configuration space, the models are often forced to fit large regions of the potential energy surface with very little data.
This hinders the efficient parameter optimization with respect to finding a good global fit.
We tackle this problem by a reweighting strategy introducing an energy-dependent weight into the loss function. We penalize configurations more than \SI{6}{eV/\mathrm{atom}} above the minimum energy with an exponentially decaying factor.
This is generally referred to as domain reweighting and prevents the model from spending too many degrees of freedom into fitting the thermodynamically less relevant high-energy domain, retaining the accurate fit of the more relevant low-energy domain.

When training the committees, we observed a strong dependence of the final weights on the initial random initialization. Consequently, the quality of the individual models varied substantially.
To avoid outlier members, we train our committee models with 10 members and after training select the $N_C=5$ that performed best.
In this fashion, we managed to establish a numerically stable training process allowing us to automate the whole AL workflow. Nevertheless, we could not entirely guarantee perfect MLIP ensemble fits resulting in a slight variance in trained model performance, which however did not significantly affect the results presented here.

\subsection{Optimizations and symmetry analysis}
Structural relaxations to categorize explored structures were performed for a maximum of 200 steps using the LBFGS optimizer implemented in the atomic simulation environment \cite{larsen_atomic_2017}. The convergence criterion for the forces was set to \SI{0.1}{eV/\angstrom}. The spacegroup analysis was performed using \texttt{spglib} \cite{togo_textttspglib_2018} with a coarse tolerance of \SI{0.3}{\angstrom}.

\section*{Author contributions}
N.U. was the main code developer, designed the workflow, generated and analyzed most of the data and composed the first version of the main manuscript. 
M.K. conducted valuable numerical experiments in a very early stage of this work and contributed to the writing process.
G.K.H.M. supervised the work, secured funding and contributed to the writing process.
All authors reviewed the manuscript.

\section{Competing interests}
All authors declare no financial or non-financial competing interests. 

\section{Data availability}

The active-learned training databases for Si, Ge and Ti generated and analysed in the current study are available on Zenodo \cite{zenodo}. 

\section{Code availability}

Our RENS implementation \texttt{JAXNEST} underlying all simulations from this study is not yet publicly available but may be made available to researchers on reasonable request from the corresponding author. 
A compatible version of \textsc{NeuralIL}, including example scripts for training and evaluation, is available on GitHub \cite{neuralil2022github}. 

\section{Acknowledgements}

This research was funded in part by the Austrian Science Fund (FWF) through 10.55776/F81 and 10.55776/COE5. For open access purposes, the authors have applied a CC BY public copyright license to any author accepted manuscript version arising from this submission.

\bibliography{NU,GM,repos}

\end{document}